# Designing large, high-efficiency, high-numerical-aperture, transmissive meta-lenses for visible light

Steven J. Byrnes,[1,2,*] Alan Lenef,[3] Francesco Aieta,[1,4] and Federico Capasso[1,5]

[1] *Harvard John A. Paulson School of Engineering and Applied Sciences, Cambridge, Massachusetts 02138, USA*
[2] *Charles Stark Draper Laboratory, Cambridge, Massachusetts 02139, USA*
[3] *OSRAM SYLVANIA, Corporate Innovation – Advanced Technology, 71 Cherry Hill Dr., Beverly Massachusetts 01915, USA*
[4] *LEIA Inc., Menlo Park, California 94025, United States*
[5] *capasso@seas.harvard.edu*
* *sbyrnes@draper.com*

**Abstract:** A metasurface lens (meta-lens) bends light using nanostructures on a flat surface. Macroscopic meta-lenses (mm- to cm-scale diameter) have been quite difficult to simulate and optimize, due to the large area, the lack of periodicity, and the billions of adjustable parameters. We describe a method for designing a large-area meta-lens that allows not only prediction of the efficiency and far-field, but also optimization of the shape and position of each individual nanostructure, with a computational cost that is almost independent of the lens size. As examples, we design three large NA=0.94 meta-lenses: One with 79% predicted efficiency for yellow light, one with dichroic properties, and one broadband lens. All have a minimum feature size of 100nm.

**OCIS codes:** (050.1965) Diffractive lenses; (050.6624) Subwavelength structures; (050.1950) Diffraction gratings; (160.3918) Metamaterials; (050.5080) Phase shift.

## 1. Introduction

In recent years, *metasurfaces* have emerged as a powerful and practical paradigm for making lenses and other optical components. To make a metasurface, one starts with a flat wafer, and decorates the surface with carefully-designed nanostructures. These alter the phase of light as it passes through or reflects, creating a new wavefront. For example, a metasurface with a hyperboloid phase profile acts as a lens with no spherical aberration [1].

For a variety of applications, including point-source collimators, laser-based microscopy, and so on, we want a macroscopically-large (millimeter- to centimeter-scale), high-numerical-aperture, polarization-insensitive, high-efficiency transmitting meta-lens for visible light. Recent work suggests that a promising approach is to make a metasurface from an array of dielectric nano-pillars [2–5]. The lens functionality would be created by judicious choice of the location and radius of each of the nano-pillars.

Thus, the lens design includes upwards of $10^9$ free parameters, or more if we permit the nano-pillars to have complicated, non-circular cross-sections. Note that there is not necessarily any periodicity to simplify this design problem—see Figs. 1(a) and 1(b). Computational optimization of photonic structures is a blossoming field [6,7]; however, in this situation, the structure may be too big to simulate directly, let alone optimize with so many free parameters. For example, a Finite-Difference Time Domain (FDTD) simulation with $10^{10}$ grid-cells requires a powerful supercomputer [8]; but this corresponds to a visible-light meta-lens with only ~1mm diameter. For many applications we may want a 100× larger area than that, and moreover we may want to run hundreds of such simulations to tweak the design and predict lens performance under different conditions. In this paper, we will show a design approach that makes full-lens simulation and pillar-by-pillar optimization feasible, even on a desktop computer.

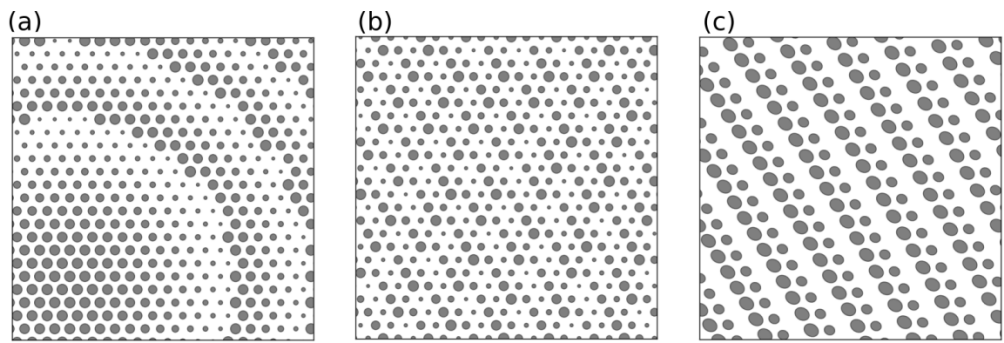

Fig. 1. (a-b) A typical dielectric metasurface, made by the traditional design method, puts nano-pillars on a regular hexagonal array, and alters their diameter to achieve the desired phase at each point. Top-down view near the center (a), and far from the center (b), showing the complex, aperiodic pattern. (c) With our modified design technique, the structure is approximately (though not perfectly) periodic in both dimensions, except near the center.

Previously, metasurfaces have been designed using a semi-analytical approach,

summarized in the next section. However, if the designs created this way are deficient in some respect, there is little room to improve them. For example, on the periphery of a high-NA visible-frequency meta-lens, this traditional approach is inherently inaccurate due to strong phase gradients, interactions between neighboring pillars, and oblique angles, as discussed in more detail below. As another example, we will show a lens that focuses yellow light, but transmits blue light undisturbed. Using pillar-by-pillar optimization, it is straightforward to make this design: we merely use an appropriate figure-of-merit. Accomplishing the same thing would be quite challenging with the traditional approach. Finally, the structures in this paper all satisfy a fabrication-related constraint: all pillar diameters, and all spacings between pillars, are at least 100nm. This opens the possibility of fabricating the meta-lenses using one-step deep-UV photolithography, which is more scalable than the typical electron-beam lithography. This constraint would be difficult to implement in the traditional approach, without excessively lowering efficiency.

The final key advantage of our design approach is that we can generate a theoretical prediction of the lens's efficiency and far-field. This has been impossible previously, because the lens is too big to simulate by orders of magnitude. (The $D_6$ symmetry of a hexagonal-grid-based metasurface, visible in Fig. 1(a), can in principle shrink the simulation volume to some extent, but only if the source is on the optical axis.) Previously, researchers have avoided this challenge. Instead, some have simulated meta-lenses with very small radii (tens of microns) [3,4], while others have simulated metasurface "cylindrical lenses"—i.e., lenses that focus in only one of the two transverse directions, but are periodic in the other direction, and therefore far easier to simulate [1,9]. These types of calculations can *qualitatively* suggest what the efficiency and far-field of a practical (millimeter-scale and round) meta-lens might be, but they do not lead to quantitative predictions that can be compared to experiments.

## 2. Traditional design method

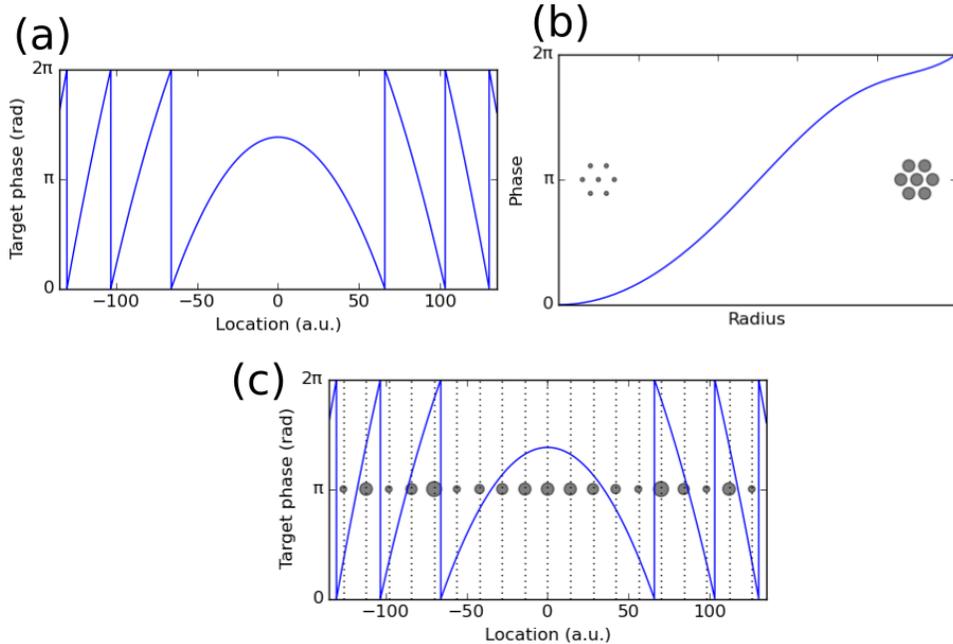

Fig. 2. The traditional metasurface design technique involves 3 steps: (a) Calculate the desired phase that the lens should impart; (b) Calculate the phase imparted by a perfectly periodic array of nano-pillars, as a function of its geometry, which hopefully spans the whole 2π range; (c) Pick a regular grid of locations for the nano-pillars (dashed lines), and choose the radius of each by combining (a) and (b) (inset circles).

The traditional meta-lens design method, as used in [2–5], as well as typical microwave transmit-arrays and reflect-arrays [10], involves three steps, summarized in Fig. 2.

In Step 1, Fig. 2(a), the target phase profile—phase as a function of position—is chosen based on the desired metasurface functionality. For example, a hyperboloid phase profile creates a spherical-aberration-free meta-lens [1], while a sawtooth phase creates a beam deflector [11] (the metasurface equivalent of a blazed grating [12]).

In Step 2, Fig. 2(b), we start by picking a motif for our metasurface that includes one or more adjustable parameters. For example, our metasurface might be a subwavelength-spaced hexagonal array of nano-pillars, each with adjustable diameter; or as another example, a far-IR metasurface might be a square array of antennas, each with adjustable shape and orientation [1,13]. Whatever the adjustable parameter is, we run a series of simulations that sweep through the possibilities. Ideally, these simulations will show a set of parameters where the transmission phase varies across the whole $0–2\pi$ range, while the transmitted intensity stays near 100%.

Then in Step 3, Fig. 2(c), we combine the results of Steps 1 and 2. We pick a regular array of subwavelength-spaced locations for the pillars, find the desired phase at each location, and then use the results of Step 2 to pick out the best nano-pillar diameter for creating that phase. At this point, the design is complete and ready to fabricate.

A shortcoming of this design approach is that each separate simulation in Step 2 is based on a periodic array of nano-pillars—or in some cases based on isolated nano-pillars [9]. But the nano-pillars in the final meta-lens will be neither periodic nor isolated—Figs. 1(a) and 1(b). Therefore it is *not* safe to assume in general that we can extrapolate the results of Step 2 to the meta-lens. For example, the reflection may be very low in the Step 2 simulations, but nevertheless higher in the Step 3 meta-lens.

More specifically, there are indeed situations where the extrapolation from Step 2 to Step 3 is likely to be quite accurate, but our target application—high-NA, visible meta-lenses—is not among them. Specifically, the extrapolation is justified in two situations. The first situation is when the nanostructures in the lens are *almost* periodic—i.e., the pillars are only slightly different than their neighbors. This happens with a low-NA lens, or for a high-NA lens if the nanostructures have *deep* subwavelength spacing. Neither is the case in our target application, except close to the lens center—Figs. 1(a) and 1(b). The second case where extrapolation is justified is if the nano-pillars do not interact much with their neighbors. This is typically true for NIR dielectric metasurfaces, but less so for visible frequency, because the nano-pillars materials available in the latter case tend to have much lower refractive index. For example, a visible-frequency metasurface might use amorphous $TiO_2$ with n~2.3, whereas a NIR metasurface might use amorphous Si with n~3.6. Therefore, the visible metasurface's pillars cannot confine light as well, leading to more interactions between neighboring pillars.

Another shortcoming of extrapolating from periodic simulations to go from Step 2 to Step 3, is that there may be angle-dependent effects. Specifically, for a metasurface to have very high efficiency, it is not enough to have the different parts of the metasurface add in phase in the desired direction. We also want the radiation pattern of the individual nano-pillars to have a substantial weight in those directions. Thus, if a nano-pillar tends to radiate upwards, it will work well near the lens center but poorly on the periphery of a high-NA lens. Ideally, we would use different nano-pillars in different parts of the meta-lens to seek out favorable radiation patterns. This would be difficult to implement in the traditional design approach, but it occurs automatically in our optimization-based designs.

## 3. Modified design method

While the traditional design leads to designs with very little symmetry or order away from the center, as in Fig. 1(b), that is not inevitable; our modified design method leads to far more ordered meta-lenses like Fig. 1(c). This order comes from exploiting two symmetries of the desired hyperboloid phase profile. The first is the circular symmetry of the lens. A hexagonal grid, as might be used in a traditional design, reduces this infinite symmetry to merely $D_6$ symmetry (visible in Fig. 1(a)). By contrast, in our designs, parts of a millimeter-scale lens might have as much as 100,000-fold rotational symmetry. Secondly, away from the center, the desired phase approaches a regular periodic sawtooth phase—see Figs. 3(a) and 3(b). The sawtooth is not *perfectly* periodic—its period gradually decreases as you move away from the lens center—but for a macroscopic lens it is *nearly* periodic. In the traditional design, the pillars are placed on a fixed-period subwavelength lattice (dashes in Fig. 3(b)) which is incommensurate with the sawtooth, so the phase is sampled differently each period of the sawtooth. This leads to complicated shifting moiré patterns as in Fig. 1(b). In our approach, by contrast, we allow the pillars to be at arbitrary locations, but we repeat the pattern of pillars every $2\pi$ period of the phase (see Fig. 3(c)). Thus the pillars can have a (nearly) periodic pattern.

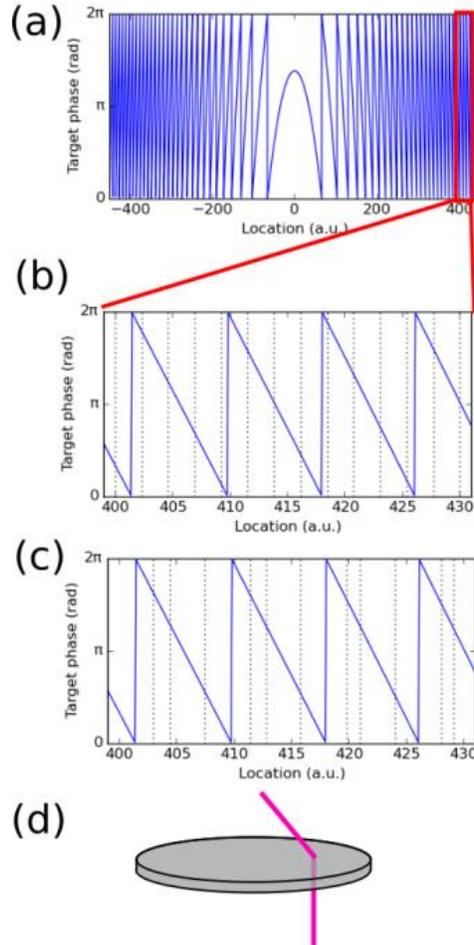

Fig. 3. (a-b) Away from the center, if we zoom in on the hyperboloid phase profile, it looks like a sawtooth—almost periodic, but with a slowly-varying period. (b) Since the traditional design approach (as in Fig. 2) places the nano-pillars on a regular grid (dashed lines), incommensurate with the phase, the resulting pattern is completely aperiodic. (c) Alternatively, we can draw pillars at arbitrary locations, but repeat the pattern every $2\pi$ phase (dashed lines). Then the overall pattern of pillars is periodic—apart from the slowly-varying period. This approach amounts to designing a meta-grating beam deflector, which is consistent with the light-bending functionality that we expect from a lens, as shown in (d).

Note that both the traditional and modified approaches involve stitching together simulations with periodic boundary conditions. The difference is that we have expanded the unit cell to include exactly one whole sawtooth of phase. This makes the periodic boundary conditions far more accurate and appropriate, especially in the region far from the lens center, when the phase is changing rapidly and (approximately) linearly. The traditional method, by contrast, works only near the center of the lens, where the phase is changing slowly compared to the separation between nano-pillars. (The nano-pillars cannot be arbitrarily small and close together, due to both fabrication constraints and neighbor-neighbor interactions.)

Our task is now to pick a periodic pattern of nano-pillars to match a sawtooth phase profile. We recognize this task as being equivalent to designing a metasurface beam deflector, i.e. the metasurface equivalent of a blazed grating [11,12]. This is consistent with our intuition about how a lens works—any small region within a large lens should act like a beam deflector (see Fig. 3(d)). In other words, we are locally approximating the hyperboloidal phase profile as a linear phase ramp. (Besides lenses, this design approach applies to any other optical element that has a locally-periodic phase profile, including beam deflectors, axicons, etc.)

Combining the circular symmetry with the phase periodicity, we get overall lens plans such as shown in Fig. 4. Each small box in Fig. 4 is to be filled in with the unit cell of a periodic metasurface beam deflector. The final resulting patterns might look like Fig. 1(c).

When designing these beam deflectors, we use simulations that have periodic boundary conditions in two of the three dimensions, with a rectangular unit cell. These types of electromagnetic simulations are exceptionally fast and practical using any number of software tools (see below), but it is important to note that it is an approximation to apply these types of simulations to our structure, for the four reasons shown in Figs. 4(i)–4(iv). First, moving from the center of the lens towards the outside, the radial dimension of the rectangular cell (Fig. 4(i)) gradually changes, breaking the periodicity, as discussed above. Second, the azimuthal dimension (Fig. 4(ii)) gradually changes for geometrical reasons. Third, relatedly, the boxes are slightly wedge-shaped rather than rectangular (Fig. 4(iii)). Fourth, we add a few "grain boundaries" (Fig. 4(iv)) where the pattern abruptly changes. This gives us more flexibility in the design, particularly controlling the azimuthal dimension of the unit cell. It is also an opportunity to discontinuously change the pattern, particularly by changing the number of nano-pillars in each unit cell.

While these four inaccuracies must be acknowledged when applying rectangular periodic simulations to our lens, all four become increasingly inconsequential as the size of the lens increases. For example, if we have a 2mm-focal-length lens designed for 500nm light, and we look 2mm away from the lens center, then it turns out that neighboring unit-cells differ in each dimension by less than 0.04%, and the two sides of the wedge-shaped region might be non-parallel by only 0.01°. We expect undesired light scattering within a wavelength or so of the grain boundaries, but if there are 5 grain boundaries on a 5mm-diameter meta-lens, then this scattering might affect only 0.1% of the area of the lens. Therefore, while conventional meta-lens simulations get more difficult as the lens size increases, our methods work better with increasing lens size—the design time and simulation time are essentially unaffected, while the accuracy increases.

As one approaches the center of the lens, the modified design method becomes progressively worse, as the approximations of Figs. 4(i)–4(iii) become less justified. Indeed, at the very center of the lens, there is no sawtooth phase whatsoever (see Fig. 3(a)). By contrast, the traditional design approach works best near the center of the lens, and worsens towards the periphery. The two are complementary; therefore we use the traditional design approach in the center of the lens, and the modified approach elsewhere. We switch between the two at ~$f$/4 distance from the center of the lens, where $f$ is the focal length, i.e. the part of the lens that bends light by 15°. In our NA~0.95 lenses, the central 1% of the lens area uses the traditional design approach, and the remaining 99% uses the modified design approach described in this paper.

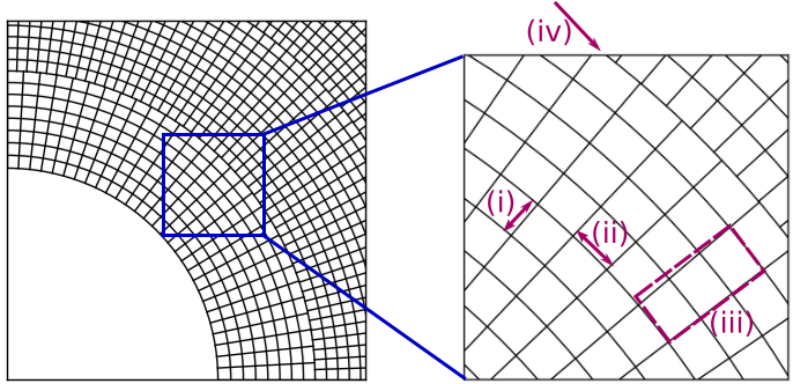

Fig. 4. The whole lens—apart from the very center—is broken into small boxes, each of which contains the unit cell for a meta-grating beam deflector (cf. Fig. 1(c)). We design these deflectors by running simulations of a rectangular cell with periodic boundary conditions. This is not exactly accurate in four respects: (i) The unit-cell size in the radial direction gradually decreases away from the center of the lens, since the periphery needs to bend light more sharply; (ii) The unit-cell size in the opposite direction gradually increases away from the center of the lens, due to geometry; (iii) Relatedly, the cells are wedge-shaped rather than exactly rectangular; and (iv) In a few places on the lens, we add "grain boundaries" where the pattern abruptly changes. All four of these inaccuracies become increasingly inconsequential as the lens size increases, and are negligible for a millimeter-scale visible light lens.

## 4. Metasurface beam deflector geometry, design, and optimizations

Using the overall approach outlined above, we have designed some specific meta-lenses. The overall geometry is shown in Fig. 5(a). For spherical-aberration-free performance, these particular devices must be used as shown, with the metasurface on the side of the substrate nearer the point source. This choice of geometry—as opposed to putting the meta-lens on the opposite side of the substrate, or even putting meta-lenses on *both* sides of the substrate (analogous to a biconvex lens)—is just used as an example. In reality, any of those three geometries might be suitable in different applications, as they have advantages and disadvantages in areas such as chromatic aberration, coma, efficiency, system integration, and so on. (These considerations are beyond the scope of this work.)

If we zoom in on the metasurface—Fig. 5(b)—we have incident light from air getting deflected to the normal in the substrate. However, some of the power will inevitably go into undesired diffraction orders (dashed lines).

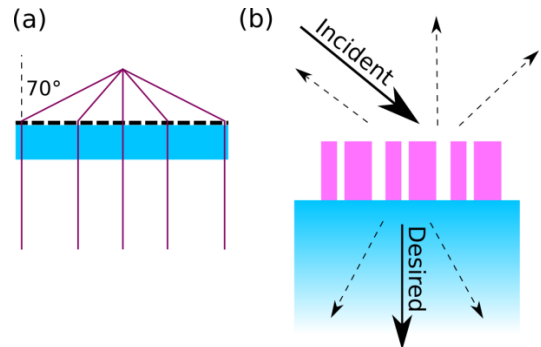

Fig. 5. Geometry of (a) the overall meta-lens, and (b) a zoomed-in section, showing the incident light, desired output, and other undesired diffraction orders (dashed lines).

(This discussion assumes for concreteness the lens is being used for collimation rather than focusing. By reciprocity, a good lens for collimation is a good lens for focusing, and vice-versa [14].)

We assume the substrate is fused silica, and the nano-pillars are amorphous $TiO_2$, using experimentally-measured refractive index data (e.g., $n_{TiO2}$=2.37 at 580nm). We found that 550nm tall pillars with elliptical cross-sections work well in simulations, and these were used exclusively. These are taller than some previous work [2,5], but still fabricable. We believe that the nano-pillars act somewhat like waveguides, with different sizes creating different effective indices, which lead to different phase shifts. With taller pillars, it is easier to create the full 0-2π range of phase shifts from a smaller variety of effective indices, leaving more flexibility to minimize reflections, tailor radiation patterns, and so forth. A more in-depth exploration is worthwhile but beyond the scope of this work.

Any part of the lens (excluding the central 1%, as discussed above) is supposed to behave like a metasurface beam deflector, which is (to a good approximation) periodic in the x and y dimension. We can simulate this structure using Rigorous Coupled-Wave Analysis (RCWA), specifically with the $S^4$ software package [15]. We cross-checked some results with FDTD (Lumerical software) and found reasonable consistency. Since RCWA simulations are quite fast, we can easily perform iterative optimization to choose a pattern of nano-pillars that has maximum efficiency for transmitting unpolarized light into the desired diffraction order.

More specifically, the figure-of-merit guiding our optimization algorithm was defined as follows. First, we calculate the power into the desired diffraction order (Fig. 5(b)), with polarization matching the source. Then, we multiply by the sine of the phase with which light enters this diffraction order, relative to the phase of the source wave, as measured at the center of the unit cell. This encourages the optimization algorithm to create the same output phase—namely π/2, the phase with maximum sine—for all the beam deflectors across the lens. The target phase π/2 is arbitrary (any other phase would have worked just as well) but we want a consistent phase across the lens in order to approach diffraction-limited performance. Finally, as the last step, we average this quantity above, diffracted power times sine of phase, over the two source polarizations (s and p), and over all relevant wavelengths. This gives the figure-of-merit which we try to maximize.

For this proof-of-principle work, we optimized this figure-of-merit using a quite simple optimization algorithm. We have a part of the lens that we are designing, which determines the length of the unit cell (Fig. 4(i)) by the grating equation. We use guessing and trial-and-error to choose the remaining aspects of the geometry—the width of the unit cell, and the number, sizes, and positions of nano-pillars. Then, we run our optimization algorithm to adjust the sizes and positions of the nano-pillars. There are 5*N* adjustable parameters, where *N* is the number of nano-pillars, since each nano-pillar has an (*x,y*) center, a major and minor axis, and an orientation angle. The optimization algorithm varies one or more of these parameters by a small amount. If the change increases the figure-of-merit, and still satisfies our fabrication-related constraints (all diameters and pillar-to-pillar separations must be >100nm), then we keep this new geometry, and otherwise we undo the change. This process is repeated many times. More specifically, we used two variants of this approach in sequence, one where we varied one parameter at a time by a fixed amount, and quit when no more changes were helpful; and another where we varied all the parameters at once by small random amounts, and quit after trying this a few hundred times. Either way, this is a very simple algorithm for seeking out the nearest local maximum, and should give similar results to the gradient-ascent method. We expect that significantly higher-efficiency meta-lens designs could be generated by using more sophisticated optimization algorithm in a larger parameter space [6,7]—for example, a wider variety of pillar cross-section shapes beyond elliptical.

Once we have found an optimized beam deflector, we can fill in one of the little rectangles of Fig. 4 with the appropriate nano-pillars. We can also immediately fill in the neighboring unit cells in the azimuthal direction, simply by slightly rotating the configuration of nano-pillars. A trickier issue is how to fill in the unit cells at different radii. Because the cell dimensions and incident light angle are gradually changing, we cannot simply copy the pattern from neighbor to neighbor in the radial direction. Instead, we need to design a collection of gradually-changing beam deflectors—a collection that will fill in the whole region between two concentric "grain boundaries" (Fig. 4(iv)). The deflectors in this collection must have different periodicities, with unit cell dimensions following the formula:

$$W \propto \tan(\sin^{-1}(\lambda/L)), \tag{1}$$

where $L$ is the length of the unit cell (in the radial direction, Fig. 4(i)), $W$ is the width (Fig. 4(ii)), and $\lambda$ is the design wavelength. This formula is derived from the grating equation and the fact that $W$ is proportional to the distance from the lens center (Fig. 4).

We build this deflector collection as follows. Each step, we increase $W$ by 1%, decrease $L$ by an amount dictated by Eq. (1), adjust the incident light angle, and then re-optimize this new deflector, using the previous deflector as an initial condition. This re-optimization process is the same as described above, but with an additional constraint that drastic changes to the geometry are forbidden—e.g., there is a limit on how much each cylinder can move. This ensures a smooth variation, and justifies the approximation that the deflector is locally almost periodic. The whole process up to here can be done before knowing the radius and focal length of the lens, since Eq. (1) does not rely on those parameters. But once we decide on the focal length, then we can finally fill in each unit cell by interpolating between the beam deflectors in our smoothly-varying deflector collection.

Once we have enough deflector collections to cover the desired range of light-bending angles (perhaps a half dozen), we have our metasurface design. More specifically, we have a design for everything but the central part of the lens, which requires the traditional metasurface design approach, as mentioned above.

Some example optimized beam deflectors will be shown in a later section (Fig. 7).

## 5. Far-field calculation

We can calculate the lens's infinite far-field using a near-field-far-field transform [16]. The basic idea is that, since the lens is locally (approximately) periodic, we can approximate the near-field at any given point by assuming that the structures near that point are *exactly* periodic, and use our RCWA deflector simulations to figure out the field amplitudes.

A subtle aspect of this calculation is what points to use for the near-field. Ordinarily, in near-field-far-field transforms, there is a trade-off governing how far the near-field points should be above the surface. If we use points many wavelengths above the surface, as in Fig. 6(a), the evanescent waves have already decayed away, and we can space the sample points by up to $\lambda/2$—the Nyquist limit for distinguishing propagating waves. However, each point is then influenced by many unit cells of the local pattern, which means our assumption of almost perfect periodicity becomes more questionable. Alternatively, we might use points closer to the surface, as in Fig. 6(b), but we need to calculate the field at a far denser array of points, or else the evanescent waves will become aliased with the propagating waves.

However, since we are simulating the metasurface beam deflectors with the RCWA method, we can avoid this trade-off altogether. RCWA directly calculates the Fourier series of the outgoing field, and only indirectly infers the real-space field. Therefore we can eliminate the evanescent Fourier components from the outgoing field in k-space, *before* calculating the field at any real-space points. With this technique, we are free to use near-space points infinitesimally above the surface, and yet still space them by up to $\lambda/2$.

Since the outgoing waves are in glass (see Fig. 5(b)), there are modes with $k_{vac}^2 < k_x^2 + k_y^2 < k_{glass}^2$ that are evanescent in air but not in glass. More specifically, the light in these modes will experience total-internal-reflection at the back surface of the substrate. In the plots and figures below we will ignore these modes, treating the energy they carry as a pure loss, similar to the reflected light—a conservative assumption.

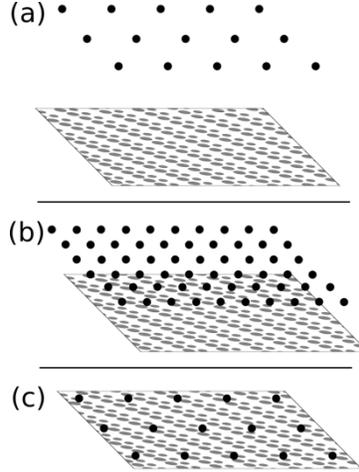

Fig. 6. Choosing near-field points for a near-field-far-field transform. (a) If we choose points a few wavelengths above the nano-pillars, the points can be spaced by $\lambda/2$. (b) If we choose points closer to the surface, we can calculate the fields more accurately, but we need far denser points to avoid evanescent wave aliasing. (c) RCWA offers the best of both worlds: We can filter out the evanescent components first, then use $\lambda/2$ spacing infinitesimally above the surface.

In more detail, the near-field calculation is implemented as follows. First, we list all the beam deflectors that are used in the lens. There might be a few hundred of them, split among 5-10 smoothly-varying deflector collections separated by "grain boundaries". (The actual lens design may use orders of magnitude more deflectors than that, but they will all be built from those few hundred by interpolation and rotation.) Then for each of these deflectors, we use RCWA to calculate the complex amplitude of each diffraction order, for a variety of incoming angles, polarizations, and wavelengths. Now, for any beam deflector on the lens, we can calculate the local outgoing field in the form $\mathbf{E}_F(\mathbf{r}) \approx \Sigma_n \mathbf{A}_n \exp(i\mathbf{k}_n \cdot \mathbf{r})$ where $\mathbf{E}_F$ denotes the outgoing electric field in the vicinity of a beam-deflector $F$; $\mathbf{r}$ is position; $\mathbf{k}_n$ are the wavevectors of the propagating diffraction orders, calculated from the local incoming field direction at $F$; and $\mathbf{A}_n$ are the complex amplitudes of each diffraction order, calculated by interpolation from the set of RCWA simulations described above. Next, we pick a regular square grid of near-field points covering the lens, and for each point we calculate what beam deflector unit cell $F$ it falls on, and finally calculate the electric field $\mathbf{E}_F(\mathbf{r})$ at that point. (We calculate $\mathbf{H}$ similarly.)

The central part of the lens, accounting for 1% of the lens area in our designs, uses the traditional design approach. The pillars are on a regular grid, but the phase gradients are sufficiently gentle that each pillar has a similar size to its neighbors. Thus we can still compute the near-field using essentially the same technique.

Once we have compiled the near-fields, we can perform a near-field-far-field transform following [16]. The main computational step involved is taking a fast Fourier transform (FFT) of the in-plane components of $\mathbf{E}$ and $\mathbf{H}$.

In our proof-of-principle implementation—the code for which is available at [17]—the entire design and simulation process can run on a desktop computer. The results shown in the next section were calculated this way over the course of a few days, with most of the time spent on the design optimization, and the rest on the far-field calculations. With a more sophisticated optimization algorithm, and better hardware, the process could be much faster. As expected, we found that if NA is held fixed while the diameter and focal length of the lens are increased or decreased, then the overall calculation time is only slightly affected, even up to millimeter- or centimeter-scale diameters.

## 6. Results

*6.1 Lens for 580nm*

For our first design, we set up the optimization algorithm to attempt to create a meta-lens for collimating 580nm light with as high efficiency as possible. We designed for a numerical aperture of 0.94, corresponding to an incident angle of 70° at the lens edge. In the design, we used six smoothly-varying collections of bean deflectors, separated by five "grain boundaries". The calculated efficiency of each individual beam deflector is shown in Fig. 7(a), with three example deflectors in Figs. 7(b)–7(d). The associated Visualization (available online) is an animation illustrating how the beam deflector geometry gradually changes as one moves away from the center of the lens. (We mention in passing that these large-angle, high-efficiency, dual-polarization beam deflectors have their own potential applications quite apart from serving as part of a meta-lens.)

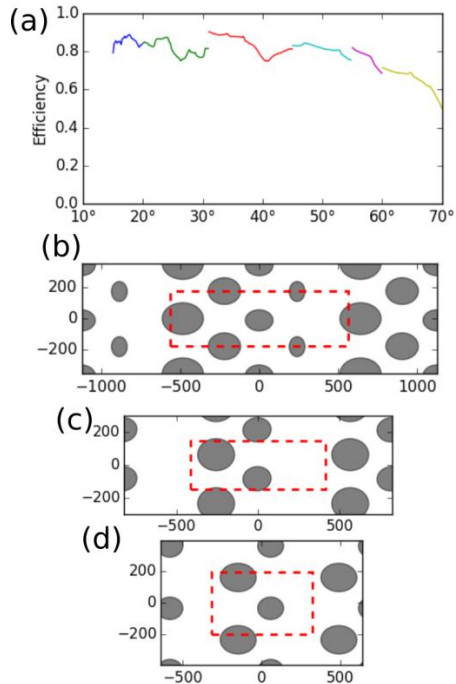

Fig. 7. (a) For all the metasurface beam deflectors comprising the 580nm meta-lens, this plot shows the calculated deflection efficiency—fraction of incident light power entering the +1 diffraction order, averaged over the two polarizations—as a function of the deflection angle, i.e. the angle between an incident ray and the +1-diffracted output ray (see Fig. 5(b)). The jumps and color changes represent "grain boundaries". (b-d) As examples, three of these deflectors are shown in a top-down view. All lengths are in nm; gray shapes are $TiO_2$ nano-pillars. (b) Bending light by 31° with 90% efficiency; (c) Bending light by 45° with 83% efficiency; (d) Bending light by 65° with 68% efficiency. The pillar widths and separations are >100nm by design; the maximum pillar major axis is ~250nm. **Visualization 1** shows an animation of a smoothly-varying deflector collection.

Next, we ran the far-field calculation for this design, assuming a lens diameter of 1mm (focal distance 200µm). For the light source, we modeled an unpolarized Lambertian source at the focal point. This entails starting with a dipole field polarized in the x-direction, scaling E and H by $(\cos\theta)^{1/2}$ to follow Lambert's law, and then calculating the far-field intensity; then doing the same for y-polarized and z-polarized dipole fields; and finally averaging all the results.

The predicted focusing efficiency (fraction of power incident upon the lens that ends up being collimated) is 79%. Unsurprisingly, this is similar to a weighted average of the deflection efficiencies in Fig. 7(a). Another 7% of the light is transmitted but scattered into other directions, while the rest either reflects off the metasurface immediately, or is transmitted but with too large a scattering angle to escape the substrate on the first pass.

Fig. 8 shows the infinite far-field for this case, showing far-field power as a function of the direction cosines ($u_x$,$u_y$) in the glass substrate. (Multiply $u_x$ and $u_y$ by 1.5 to get the direction cosines in air, i.e. after refracting at the back of the substrate. The plot shows power per area in the $u_x$-$u_y$ plane; divide it by $u_z$ to get power per steradian.) We see that the light is collimated into a very narrow cone of angles, close to the diffraction limit. The log scale plot, Figs. 8(c) and 8(d), makes it possible to see some Airy rings, and also to see the faint light spread into other angles. Note that Fig. 8 shows very little power outside the circle $(u_x^2+u_y^2)^{1/2} \sim 0.65$. This is not a real effect; it is a consequence of the fact that, for simplicity, we did not calculate the diffraction orders that are too tilted to escape the substrate. (This light is properly taken into account in the collimation efficiency calculations, but omitted from the plots in Fig. 8.)

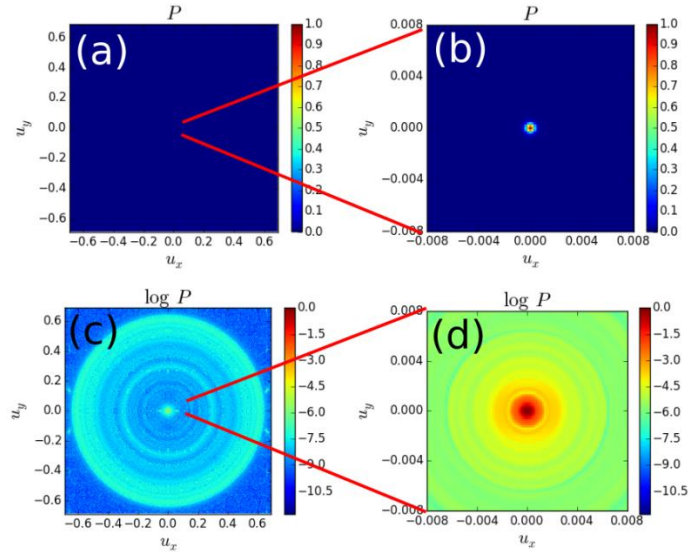

Fig. 8. Calculated far-field power (in arbitrary units) for the 580nm lens. (a-b) Linear scale, (c-d) Log scale. The plots on the right are zoomed in towards the normal direction. ($u_x$, $u_y$) are the direction cosines in the glass substrate. The power is specifically "power per area in the $u_x$-$u_y$ plane", which near the center is similar to power per steradian.

We additionally did some calculations for scaled-up and scaled-down lenses with the same NA. We found that the lens size and focal distance made little difference to the collimation efficiency or any other results, except for the predictable narrowing of the output beam angle as the lens size increased, consistent with our expectations from the diffraction limit.

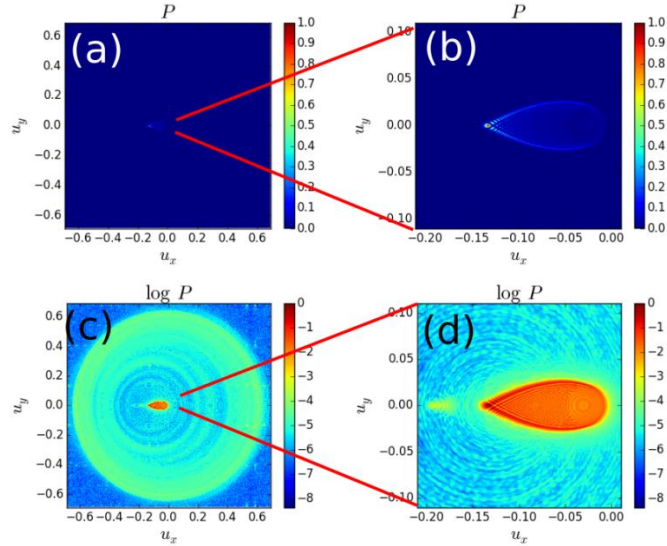

Fig. 9. Calculated far-field power for the 580nm lens, with the source moved off axis from the focal point by 1/5[th] of the focal distance. (a-b) Linear scale, (c-d) Log scale. We see a classic coma aberration.

Next, we moved the source off-axis by 1/5[th] of the focal distance. The far-field is shown in Fig. 9. We see quite clearly the coma aberration, as expected from any flat lens, particularly with such a high NA [18]. But encouragingly, the efficiency is not appreciably lowered. Indeed, the coma (teardrop-shaped) region of the far-field in Fig. 9 contains about 75% of the light power incident upon the lens, only slightly lower than the 79% for an on-axis source.

### 6.2 Lens for focusing 580nm and transmitting 450nm

Our second lens was designed to focus 580nm light with the same 0.94 numerical aperture and 1mm diameter as above, while simultaneously allowing normally-incident 450nm light to pass through unperturbed. This is an example of the kind of dichroic functionality which is difficult to achieve using conventional optical components, and it also serves as a good test of our design approach.

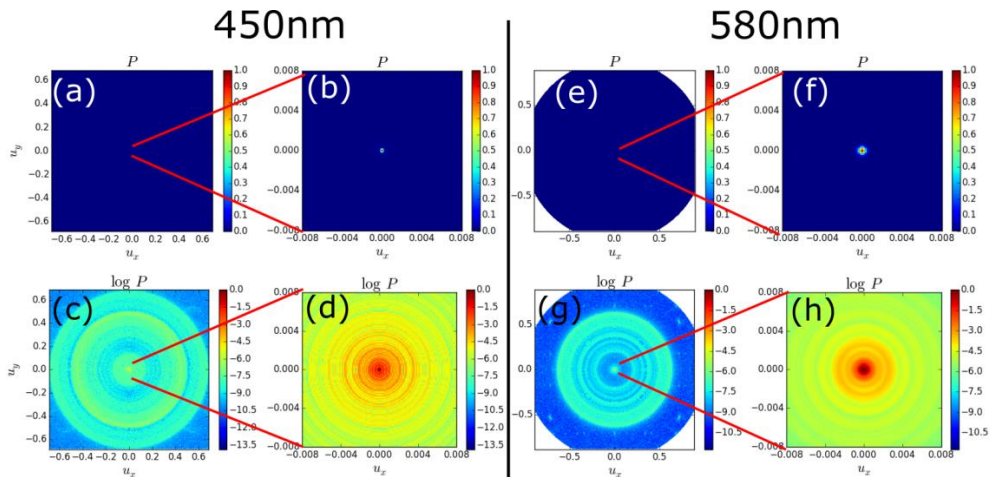

Fig. 10. Calculated far-field power for the 580nm-focusing, 450nm-transmitting lens, where the light source is (a-d) a normally-incident 450nm plane wave; (e-h) 580nm light coming from the focal point. Labels are as in Fig. 9.

When we adjusted our optimization algorithm for this goal—maximizing the –1 diffraction order at 580nm and the 0 diffraction order at 450nm—we found that the resulting lens focused 68% of unpolarized Lambertian 580nm light coming from the focal point, and transmitted 58% of a 450nm normally-incident plane wave unperturbed. An additional 18% of the 450nm light and 16% of the 580nm light was transmitted, but into off-normal directions. The far-field power distributions are shown in Fig. 10.

*6.3 Broadband lens for 500nm-650nm*

For the third lens, we tried to efficiently focus light across a broad bandwidth from 500–650nm. Our goal was *not* to create an *achromatic* meta-lens [9,19,20]: The different wavelengths will have different focal lengths—following the universal formula for diffractive optics (zone plates etc.)—as well as spherical aberrations. Specifically, all wavelengths see the same phase gradient, but that does not correspond to the same deflection angle. Nevertheless, we can try to design a metasurface which is *efficient* across a broad bandwidth—low reflection and low scattering—even if there are chromatic and other aberrations. Such a lens may be useful for non-imaging applications, or even for imaging if the aberrations can be corrected elsewhere in the system.

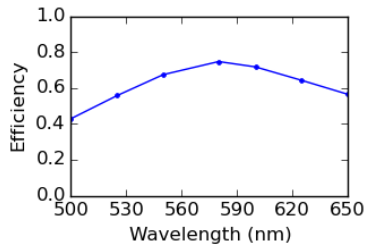

Fig. 11. Efficiency vs. wavelength for our broadband lens. As usual, efficiency is defined as the fraction of power incident upon the lens that winds up in the (possibly blurry) focal spot.

Therefore we set our optimization algorithm to maximize how well light at 500nm, 580nm, and 650nm is deflected into the –1 diffraction order, averaging over these three cases and both polarizations. The far-field calculations showed the expected focal length shifts and spherical aberrations, and we found the efficiencies shown in Fig. 11. Although we only checked three wavelengths during the optimization process, the designs worked reasonably well over the whole 500-650nm range, with highest efficiency (75%) at the center of the range, and lower efficiency towards the edges.

**7. Conclusion**

We have presented a method for designing meta-lenses with various advantages, particularly for high-NA visible-light lenses where the traditional design approach is problematic. This method makes it feasible to apply electromagnetic optimization algorithms to every nano-pillar even on a millimeter-scale lens, and enables quantitative calculations of the infinite far-field for any arbitrary source, even off the lens axis. We found that, even using a simple optimization algorithm in a restricted parameter space, we could design meta-lenses with efficiency as high as 79% at 0.94 numerical aperture in air, enforce lithography-related constraints, and create broadband or multi-functional designs, and moreover perform both the design and the simulations for millimeter-scale lenses on an ordinary desktop computer.

**Acknowledgments**

This work was supported by Osram as a Technology Cooperation (TCO) external funded

project (award A26138). SJB and FC also acknowledge financial support from Charles Stark Draper Laboratory (award SC001-0000000827).